\documentclass{PoS}

\usepackage[utf8x]{inputenc}

\usepackage{mciteplus}

\newcommand{\gev}{\ \textrm{GeV}}

\newcommand{\lqcd}{\Lambda_{\mathrm{QCD}}}
\newcommand{\as}{\alpha_{\mathrm{s}}}

\title{Small-$x$ physics at the LHeC}

\ShortTitle{Small-$x$ physics at the LHeC}

\author{\speaker{Heikki Mäntysaari}\\
        Department of Physics, University of Jyväskylä, P.O. Box 35, 40014 University of Jyväskylä, Finland\\
        E-mail: \email{heikki.mantysaari@jyu.fi}}

\abstract{The Large Hadron-electron Collider LHeC is a proposed upgrade of the LHC. It would add an electron beam to the LHC, and make it possible to study electron-proton and electron-nucleus collisions at very high energies. We present some of the highlights of the LHeC physics program related to the studies of partonic structure of protons and nuclei, and to the non-linear QCD phenomena  visible at small $x$.}

\FullConference{International Conference on Hard and Electromagnetic Probes of High-Energy Nuclear Collisions\\
		30 September - 5 October 2018\\
		Aix-Les-Bains, Savoie, France}

\begin{document}

\section{Introduction}

Deep inelastic scattering is a powerful tool to access the internal structure of protons and nuclei. Thanks to its pointlike structure, electron is a clean source of virtual photons that can resolve the hadron structure at very small distance scales. This has been seen in particular at HERA, where electron-proton collisions made it possible to measure the partonic structure of proton at an unprecedented accuracy~\cite{Abramowicz:2015mha}, and discover the rapid rise of the gluon distribution towards a small momentum fraction $x$.

The rapidly increasing gluon density suggests that one eventually enters  the regime where the gluon densities become of the order of the inverse coupling. This is a novel region of the ``cold QCD phase diagram'' where physics is non-linear, but perturbative treatments can still be possible if the scale at which this happens is large compared with $\lqcd$. As this emergent scale, known as the saturation scale $Q_s^2$, increases towards small $x$, large center of mass energies are required to accurately probe the QCD dynamics in this part of the phase space. The proposed LHeC collider cover a wide kinematical range down to $x\sim 10^{-6}$ in the perturbative range $Q^2\gtrsim 1\gev^2$ making it an ideal machine to study small-$x$ physics~\cite{AbelleiraFernandez:2012cc}. In addition to LHeC, there are also proposals to build a lower center-of-mass energy electron-ion collider in the United States~\cite{Accardi:2012qut,Aschenauer:2017jsk}.

Theoretically, the small-$x$ region of QCD can be described in the effective field theory known as the Color Glass Condensate (CGC). In CGC, one can derive perturbative evolution equations such as BK or JIMWLK, that describe the evolution of the small-$x$ gluon fields in the non-linear regime. Probing these nonlinearities at the LHeC is crucial to test the saturation picture and to understand in detail the properties of the QCD matter at extreme densities

\section{Structure functions}

Measurements of the proton and nuclear structure functions $F_2$ and $F_L$ are crucial input for extractions of parton distribution functions. Even in case of protons, the gluon distribution at small $x$ is not accurately constrained.  In addition to the extended kinematical reach compared to HERA, the possibility to precisely measure  the longitudinal structure function $F_L$ and charm contribution to $F_2$ provide valuable input to proton PDF fits. This will, in turn, have significant effect on many studies at the LHC where one of the largest uncertainties in some cases originates from the uncertainty in the proton PDFs. 

The large kinematical reach down to small $x$ is also crucial to study the possible breakdown of the fixed order perturbative calculations in the collinear factorization framework.  When gluon densities become of the same order as the inverse coupling, one effectively has to resum terms of the order $\as \ln 1/x$ to all orders. In the CGC picture the non-linear evolution equations resum these contributions. In the collinear factorization picture, it has been recently shown in Ref.~\cite{Ball:2017otu} how this resummation (which does not include non-linear contributions) improves the description of the HERA $F_L$ data compared to the nexto-to-next-to-leading order calculations. However, the effect is largest at small $x$ where HERA kinematical coverage runs out, and the large uncertainties in the HERA data make it difficult to draw precise conclusions. 

For nuclei, there has never been an experiment to study deep inelastic scattering off nuclei at collider energies. Consequently, the nuclear parton distribution functions are much less precisely constrained, and future nuclear DIS experiments will bring the field of nuclear PDFs to a completely new level of precision. Here, it is interesting to study how the partonic structure of nuclei differs from an incoherent superposition of $A$ nucleons. Especially at small $x$ we expect to see relatively strong nuclear suppression in the nuclear parton distribution functions at $x \lesssim 10^{-3}$ not accessed in previous DIS experiments. LHeC will have a large impact on nuclear PDFs in this unexplored kinematical regime~\cite{Paukkunen:2017phq}.

\begin{figure*}[tb]
 \begin{minipage}{0.48\textwidth}
\centering
		\includegraphics[width=\textwidth]{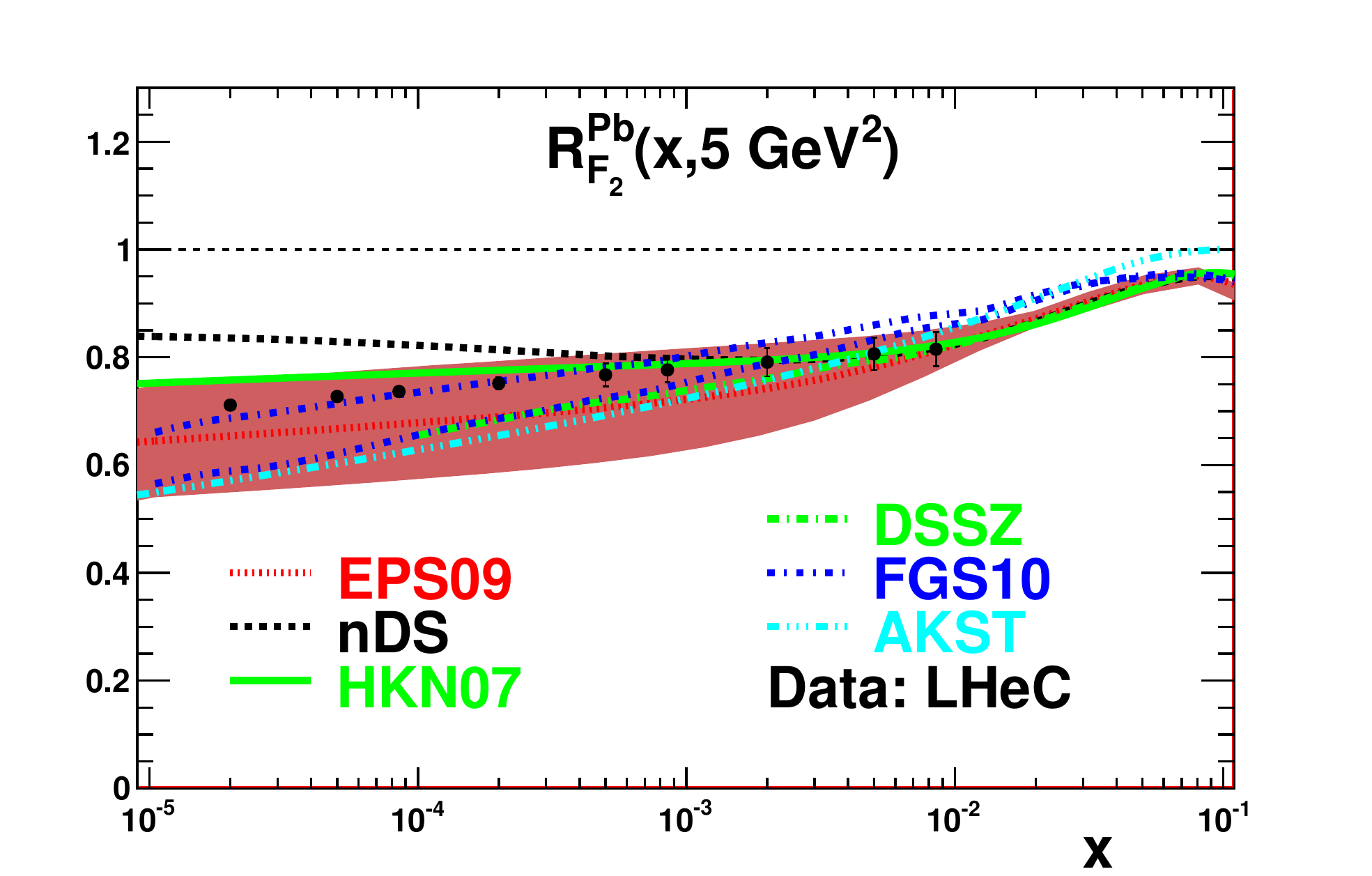} 
				\caption{LHeC pseudodata for the nuclear suppression factor at $Q^2=5\gev^2$. Figure from Ref.~\cite{AbelleiraFernandez:2012cc}.}
		\label{fig:reA}
\end{minipage}
\quad
\centering
\begin{minipage}{0.48\textwidth}
		\includegraphics[width=\textwidth]{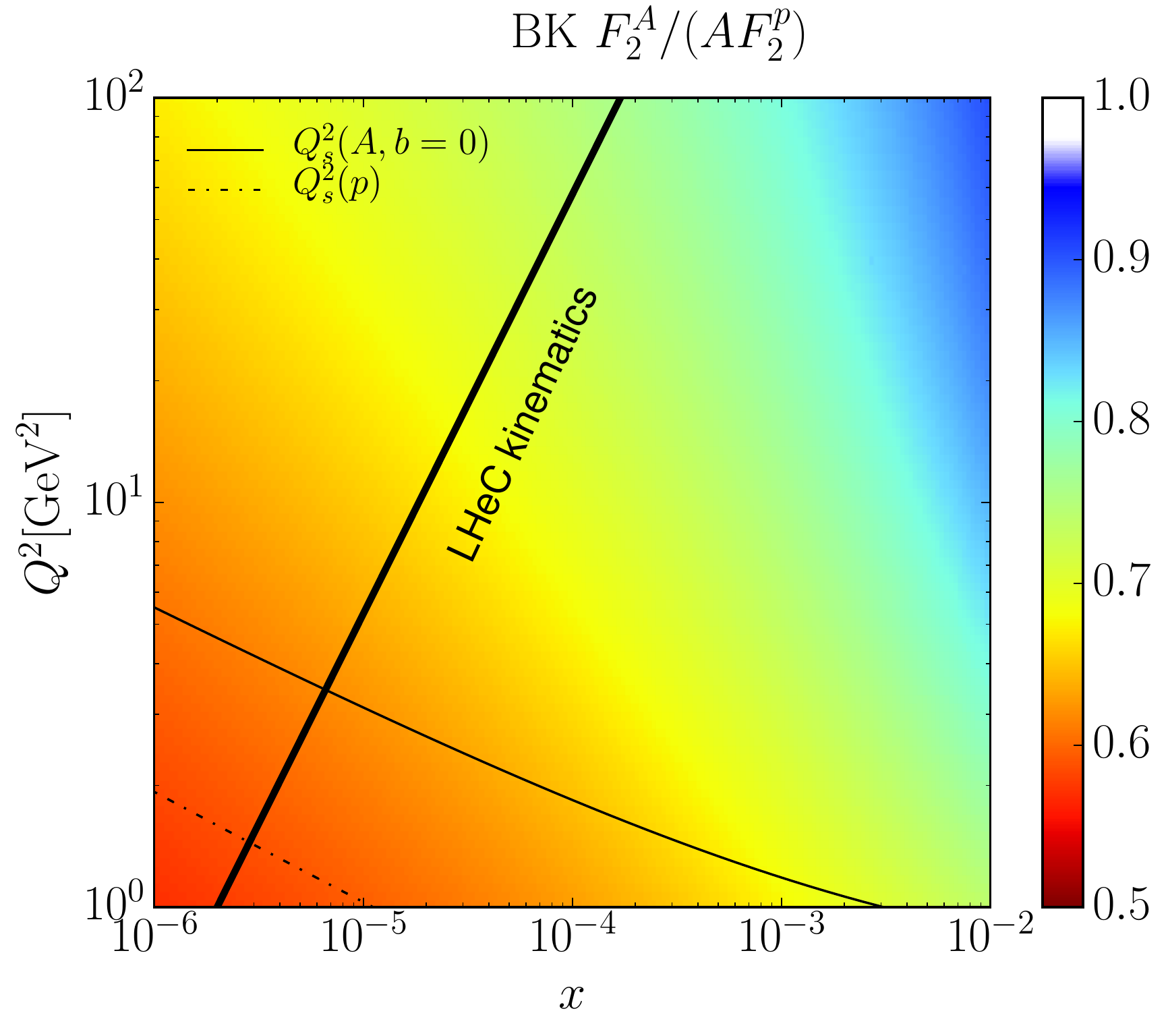} 
						\caption{Nuclear suppression function for the $F_2$ structure function in the LHeC kinmeatics.}
		\label{fig:sigmar}
\end{minipage}

\end{figure*}

Perhaps the simplest observable to study nuclear effects is to measure the structure function ratios. In the absense of  nuclear effects, the total virtual photon-nucleus cross section off the nucleus is $A$ times the corresponding cross section off the proton, and the structure function ratio $R=F_2/(A F_p)=1$. In Fig.~\ref{fig:reA} we show the LHeC pseudodata for the structure function ratio at $Q^2=5\gev^2$. The extreme precision accessible at the LHeC is clearly visible. In the same plot, comparison to some nuclear PDF predictions are shown. However, one should keep in mind that the uncertainties of the nuclear PDFs at $x \lesssim 10^{-3}$ are underestimated, as there is a large dependence on how the $x$ dependence is parametrised in the region where no data is included in the PDF fits.

In the CGC framework, the Bjorken-$x$ dependence can be computed by solving the perturbative evolution equations with initial condition fitted e.g. to the HERA data (but the $Q^2$ evolution from the DGLAP equations is not fully included) and generalized for nuclei. In Fig.~\ref{fig:sigmar} we show the predictions for the $x$ and $Q^2$ dependence of the nuclear suppression factor $R$ from the CGC calculation based on Ref.~\cite{Lappi:2013zma}. Within the LHeC kinematical coverage one can probe the transition from deep inside the saturation region where $R\sim 0.5$ up to the dilute $R=1$ region.

 \begin{figure*}[tb]
\centering
\begin{minipage}{0.48\textwidth}
		\includegraphics[width=\textwidth]{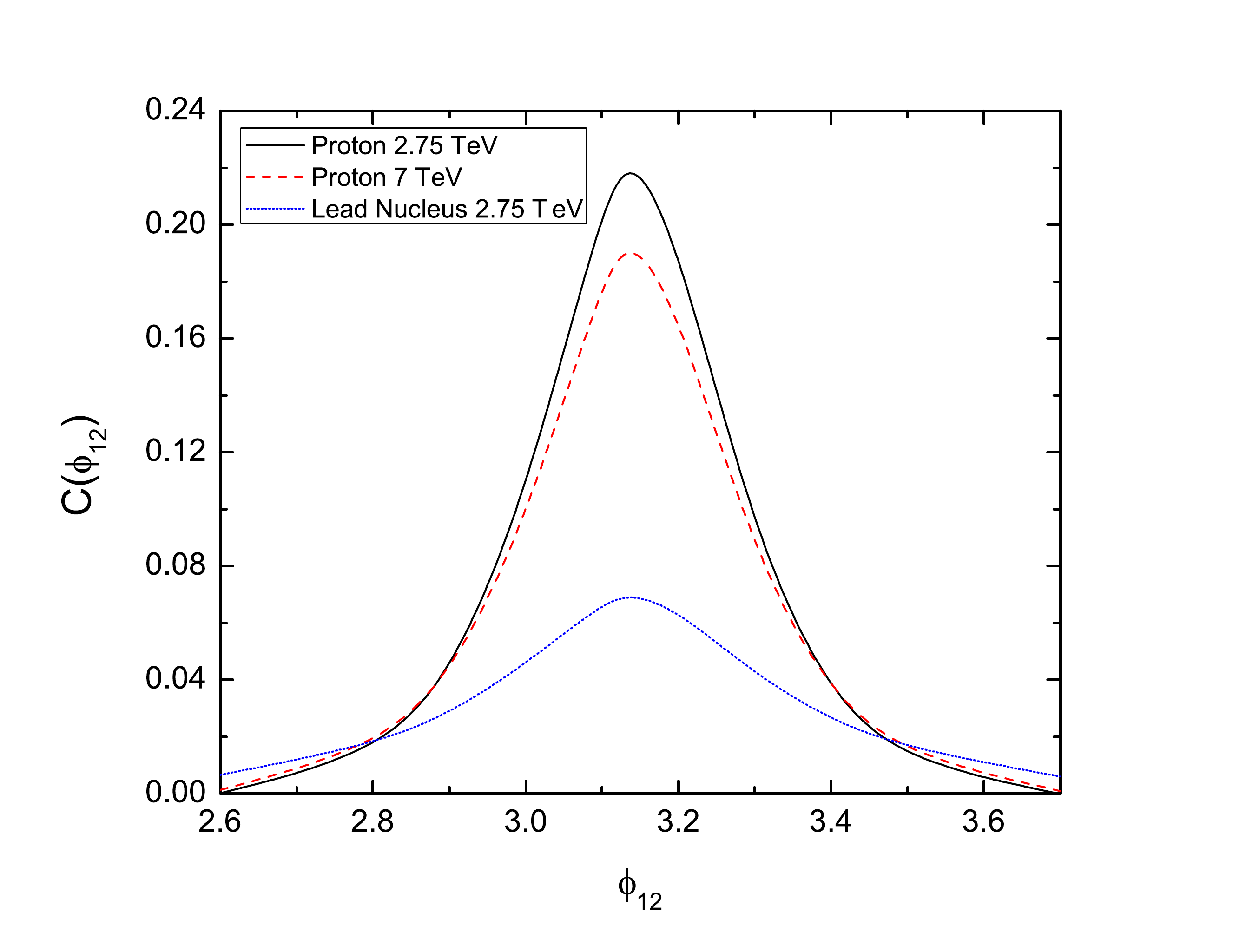} 
				\caption{ Azimuthal  correlation of two pions in $ep$ and $eA$ collisions. Figure from Ref.~\cite{AbelleiraFernandez:2012cc}.	}
		\label{fig:dihadron}
\end{minipage}
\quad
\begin{minipage}{0.48\textwidth}
\centering
		\includegraphics[width=\textwidth]{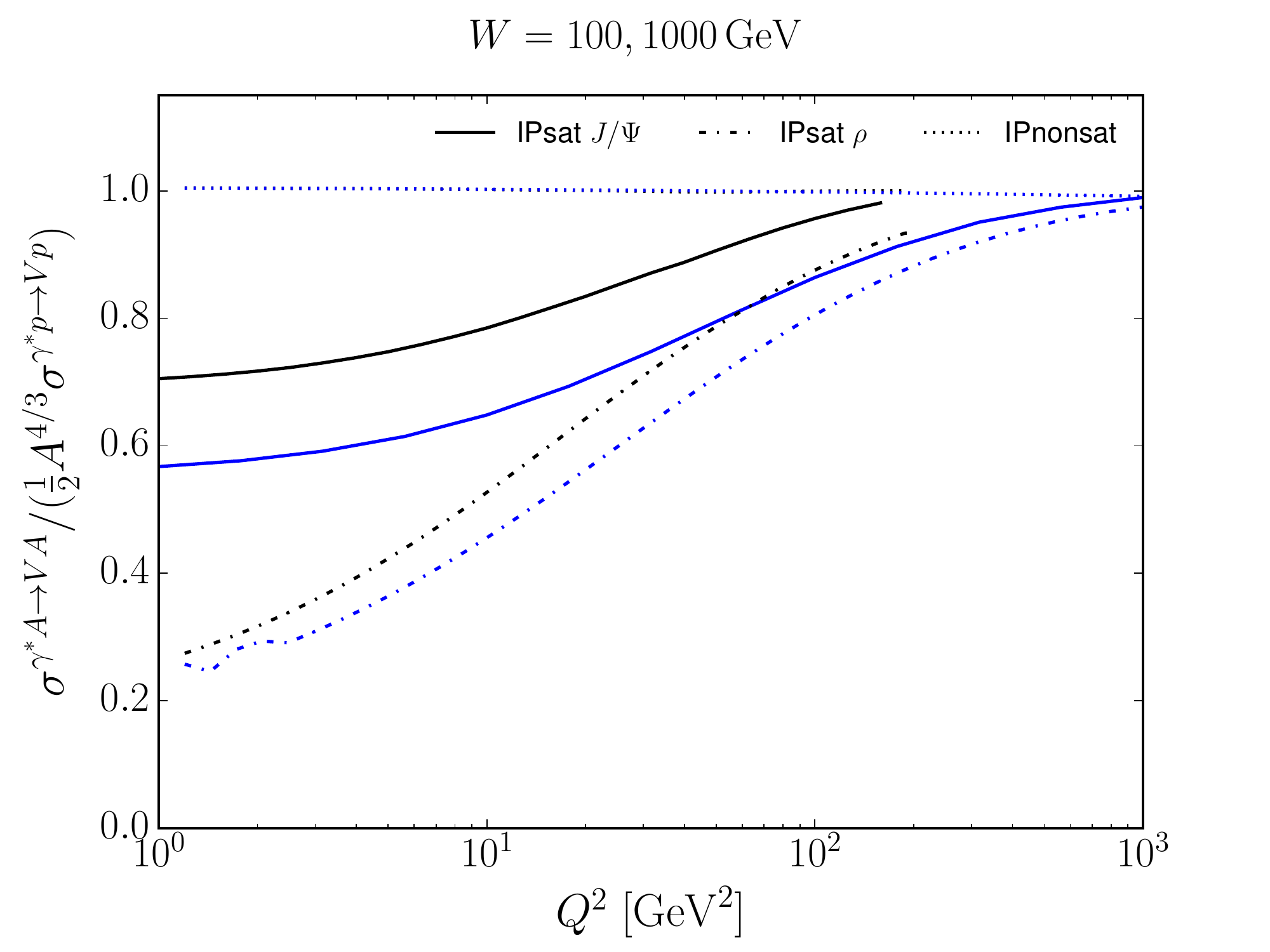} 
				\caption{Nucelar suppression factor for the exclusive vector meson production. Figure from Ref.~\cite{Mantysaari:2018nng}.}
				
		\label{fig:vm_suppression}
\end{minipage}
\end{figure*}

\section{Beyond inclusive observables}

More detailed information about the small-$x$ dynamics can be obtained by studying more differential observables. One promising quantity sensitive to the saturation effects is the angular decorrelation of forward hadrons. Here, the idea is that the virtual photon emitted from the electron fluctuates to a quark-antiquark dipole, and the two quarks are initially back-to-back. When the quarks scatter off the target, they receive a transverse momentum kick which is of the order of the saturation scale of the target, and eventually fragment in observable hadrons. As the saturation scale of the nucleus is enhanced by $\sim A^{1/3}$ compared to the proton case, the kick which washes out the back-to-back correlation is much larger with nuclear targets. 

The dihadron angular decorrelation is shown in Fig.~\ref{fig:dihadron}, where the two-pion production cross section is calculated as a function of the azimuthal angle difference. When the proton is replaced by a heavy nucleus, the back-to-back peak almost disappears, and the effect is larger at larger center-of-mass energies where one probes smaller $x$ and larger saturation scales. Similar effects have been seen in deuteron-gold collisions at RHIC~\cite{Adare:2013piz} and explained in the saturation model calculations~\cite{Lappi:2012nh}.

In addition to inclusive observables, exclusive (or diffractive) processes provide a powerful probe for the small-$x$ dynamics. In exclusive vector meson production one requires that there is no exchange of color charge between the target and the produced vector meson. This requires, at leading order, an exchange of two gluons, and the cross section becomes sensitive to the \emph{squared} gluon density~\cite{Ryskin:1992ui}. Thus, the non-linear effects should also be enhanced. In addition, as it is possible to measure total momentum transfer in exclusive processes it becomes possible to access the transverse structure and structure fluctuations of the target, as the impact parameter is Fourier conjugate to the momentum transfer (see e.g. Ref.~\cite{Mantysaari:2016ykx} and references therein). 

In Fig.~\ref{fig:vm_suppression} we show the nuclear suppression factor for the vector meson production. The nucleus-to-proton ratio is normalised such that the ratio is approximately $1$ in case the nucleus is an incoherent superposition of $A$ nucleons, as is the case in the ``IPnonsat'' parametrization shown in the figure for comparison. As can be seen, the light mesons are much more heavily suppressed than heavier $J/\Psi$, and with large center-of-mass energy one can probe the transition to the dilute region at high $Q^2$ where non-linear effects are suppressed. Note that both non-linear ``IPsat'' and linear ``IPnonsat'' parametrization provide an excellent description of both inclusive structure function and exclusive vector meson production data from HERA~\cite{Mantysaari:2018nng}.

In addition to nucelar suppression factors, one can study different scaling laws to see the onset of gluon saturation, and the disappearance of the saturation effects when moving into the dilute region. As discussed in Ref.~\cite{Mantysaari:2017slo}, a large $Q^2$ lever arm is necessary for these studies. Additionally, studying the nuclear mass number dependence of the cross section provides a good access to saturation dynamics.

It is also possible to study inclusive diffraction, where one produces a system with mass $M_X^2$ requiring the presence of a rapidity gap between the produced system and the target. In this case, the saturation model predicts~\cite{Kowalski:2008sa} that there is both nuclear suppression (at small $M_X$) and nuclear enhancement (at large $M_X^2$ where one has to include $\gamma \to q\bar q g$ splitting and consider a scattering of a $q\bar q g$ dipole), see also Ref.~\cite{Aschenauer:2017jsk}. The fraction of diffractive events should also be enhanced with nuclear targets.

\section{Conclusions}
The proposed LHeC collider has large potential to accurately probe the non-linear QCD dynamics at small Bjorken-$x$. There is a vast amount of interesting studies to be performed in such a collider. Here, we have only discussed a subset of studies that are most relevant for the small-$x$ physics. Uncovering how quantum field theory behaves in a strongly non-linear regime is an intriguing task on its own. On top of that, the LHeC studies will also have a large impact on searches at the LHC thanks to, for example, improved description of the parton distribution functions.

\subsection*{Acknowledgments}
This work was supported by the Academy of Finland, project 314764, and by the European Research Council, Grant ERC-2015-CoG-681707.

\bibliographystyle{h-physrev4mod2}
\bibliography{../../refs}

\end{document}